\documentclass[epj]{svjour}

\usepackage[OT4]{fontenc}
\usepackage[utf-8]{inputenc}
\usepackage{graphicx}
\usepackage{amsmath}
\usepackage{amsfonts}

\title{Collective firm bankruptcies and phase transition in rating dynamics}
\titlerunning{Collective firm bankruptcies}

\author{Paweł Sieczka \thanks{e-mail: psieczka@if.pw.edu.pl} and Janusz A. Hołyst \thanks{e-mail: jholyst@if.pw.edu.pl}}
\institute{Faculty of Physics, Center of Excellence for Complex Systems Research, Warsaw University of Technology, \\
Koszykowa 75, PL-00-662 Warsaw, Poland}

\date{Received: date }
\begin{document}

\abstract{
We present a simple model of firm rating evolution. We consider two sources of defaults: individual dynamics of economic development and Potts-like interactions between firms. We show that such a defined model leads to phase transition, which results in collective defaults. The existence of the collective phase depends on the mean interaction strength. For small interaction strength parameters, there are many independent bankruptcies of individual companies. For large parameters, there are giant collective defaults of firm clusters. In the case when the individual firm dynamics favors dumping of rating changes, there is an optimal strength of the firm's interactions from the systemic risk point of view.
}

\maketitle

\section{Introduction}

Banking activity is exposed to various risks. Understanding and quantifying these risks is crucial for bank management as well as for stability of the whole economy. In this paper, we focus on the bankruptcy phenomenon that is one of the factors generating credit risk for partners of a bankrupted firm. 

Credit risk at the customer level is caused by a possibility of a debtor's non-payment of an obligation.
 At the portfolio level, economic interactions play an important role. These interactions are responsible for collective defaults. 

In the economy, firms do not act as independent market participants, but are involved in complicated relations, which makes the economy an evolving, complex system.  These interactions have different sources such as: trading partnership, market competition, or simply dependence on common factors, leading to correlated defaults.

The study of default mechanisms is important for at least two reasons. The first is practical. Banks estimate probability of default (PD) to rate clients credibility and to evaluate expected losses (EL). Correlations of defaults influence unexpected losses (UL) for which banks are o\-bli\-ged to maintain regulatory capital \cite{Basel}. The second reason is theoretical. In order to work out effective rules for risk management and adequate regulations, we need to understand the mechanism of default.

Mutual correlation of credit worthiness can be modelled by the introduction of one or many risk factors which influence a company's asset value. A version of a structural model for credit risk, in which risk factors evolve according to a jump-diffusion process, was presented in Ref. \cite{Schafer}. 

Different mechanisms generating correlations are presented in works that investigate direct interactions between economical partners. Allen and Gale \cite{Allen} studied the phenomena of financial contagion. They analyzed an equilibrium model in  which credit contagion can propagate between nodes connected with borrowing-lending relations. Eisenberg and Noe \cite{Eisenberg} proposed a general model of a clearing mechanism for complex financial systems. It can be used to simulate propagation of financial distress in a network of both firms and banks.

Models of production network, in which nodes representing firms  are connected by supplier-customer relations, were introduced in \cite{Battiston,Weisbuch}. This framework allows the investigation of avalanches of bankruptcies caused by local production or delivery failures. Default cascades were also studied in \cite{Horst}. Financial distress of a single firm could propagate through the whole system, causing a cascade of downgrades and defaults due to the interactive structure of credit ratings. The problem of modelling correlated defaults due to dependencies between debtors in a credit portfolio  was also investigated in \cite{Hatchett,Egloff}.

An interbank market itself can also be a source of a financial contagion. Distress can propagate within a banking system triggering off avalanches of failures \cite{Iori,Iori2}. A collective bankruptcies model of banking networks based on random directed percolation was introduced in \cite{Aleksiejuk1,Aleksiejuk2}. This model leads to a power-law distribution of contagion sizes depending on the network dimensionality.

A default can be defined as a state of an obligor when: \textit{a bank considers that the obligor is unlikely to repay its obligation without recourse by the bank to actions such as realising security; or the obligor is past due for more than 90 days on any material credit obligation to the bank} \cite{Basel}. In this paper, we identify default with bankruptcy and use both terms equivalently.

We present a simple model of rating dynamics. We consider two sources of rating transition: the individual dynamics of a firm and its economic interactions with others. Our studies show that in such a defined environment, the phenomenon of phase transition is present. This effect has an influence on the number of defaults and on the risk of a credit portfolio. Moreover, it is responsible for collective bankruptcies, which can be observed for a certain set of parameters.

\section{The model}

\begin{figure}
\includegraphics[angle=-90, scale=0.5]{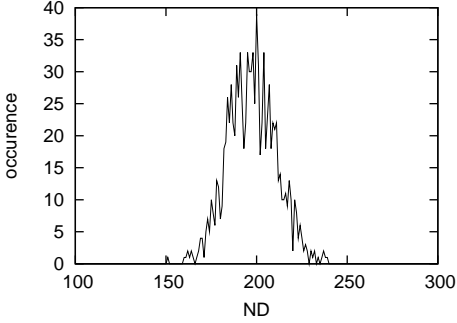}
 \caption{Distribution of $ND$ for $J_0=0.0001$, $\sigma=0.001$.}
 \label{fig:rp}
\end{figure}
We assume that a firm's financial condition can be described by a single variable $R$ taking discrete values: 0, 1, ..., $R_{max}$. This variable corresponds to a rating class that can be assigned to a firm. Each state of $R$ is related to a specific default rate, such that it grows with decreasing value of $R$. By $R=0$ we understand the state of default.

Variable $R$ evolves in time according to:
\begin{equation}
 R(t)=R(t-1)+s(t),
\label{eq:R}
\end{equation}
where $s(t)$ is a stochastic variable and $s(t)=-1, 0, 1$. This means that the rating can change no more than one class during a time step, i.e., $|R(t)-R(t-1)|\leq 1$. 

For a set of N firms, we define the conditional probability for a variable $s_i$, $i=1, 2, ..., N$:

\begin{equation}
\begin{split}
\label{eq:Ps}
 &P(s_i\setminus s_1, ..., s_{i-1}, s_{i+1}, ..., s_N, R_i)= \\ 
&=\frac{1}{Z}\exp(\sum_{j\neq i}J_{ij}\delta(s_i,s_j)+ f(R_i, s_i)).
\end{split}
\end{equation}
In the above equation, we used the following convention: a variable that stands on the left side of $\setminus$ depends on $t$, while the variables standing on the right hand side depend on $t-1$.

For a firm, the probability of $s$ taking a value (at time $t+1$) is conditional on the value of $R$ and to all the other firms' values of $s$ (at time $t$).  The states of the neighbors contribute via term $J_{ij}\delta(s_i,s_j)$, where $\delta(s_i,s_j)$ is a Kronecker delta and $J_{ij}$ models the interaction between firm \textit{i} and \textit{j}. The term $f(R_i, s_i)$ is responsible for rating transitions resulting from a specific situation of the firm. Proper normalization is given by term $Z$.

Interactions, modelled by matrix $J_{ij}$, have their source in economic connections between firms. Interactions can be realized by a positive ($J_{ij}>0$) or a negative ($J_{ij}<0$) coupling. A positive coupling appears when two firms cooperate in a general sense. Such a cooperation would mean the existence of a supplier-customer connection itself or together with trade credit involvement. If the financial condition of a given node gets worse, it may reduce its turnover and thus, diminish a profit of its suppliers or customers. The second result would be delays in payments of trade credit obligations, which affects financial solvability of the positively coupled counterparties. In both cases a change of financial condition of a node $i$ can induce a change in the same direction of a condition of a node $j$  if $J_{ij}>0$. The situation when two firms compete with each other is described by $J_{ij}<0$. Worse situation of a node $i$ can be the chance for its competitors to enlarge their share in the market.  

The values $J_{ij}$ in our model are generated from Gaussian distribution with mean $J_0$ and variance $\sigma^2_J$:
\begin{equation}
 P(J_{ij})=\frac{1}{\sqrt{2\pi \sigma_J^2}}\exp(-\frac{(J_{ij}-J_0)^2}{2\sigma_J^2}).
\end{equation}

A firm can also change its rating class as a result of  individual dynamics.  This effect is modelled by the term $f(R_i, s_i)$. The probability of a change of  class depends on the present state of $R_i$. 

Because the space of $R$-s is restriced and the bankrupted firm never recovers, special constraints should be imposed. The barrier in $R=0$ should be absorbing and the barrier in $R=R_{max+1}$ should be reflecting. This is equivalent to the condition:
\begin{equation}
 R(t)=R(t-1)+s(t)+\eta(R(t-1), s(t)),
\end{equation}
where: (\textit{i}) $\eta(R,s)=-s$ if $R=0$, (\textit{ii}) $\eta(R,s)=-1$ if $R=R_{max}$ and $s=1$, (\textit{iii}) $\eta(R,s)=0$ in the remaining cases. Thus, for $R(t-1)=0$ the rating variable $R(t)$ will always be equal to $0$ and for $R(t-1)=R_{max}$ $R(t)$ will not be greater than $R_{max}$.

\section{Simulations}

\begin{figure*}
 \includegraphics[ angle=-90, scale=0.5]{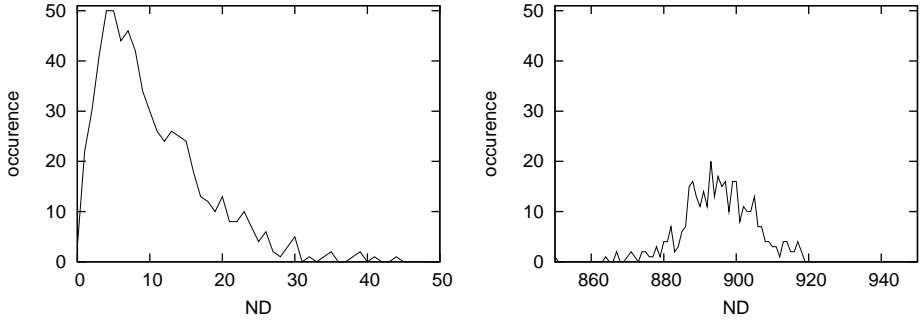}
 \caption{Distribution of $ND$ for $J_0=0.02$ $\sigma=0.001$. For better visibility the distribution was divided into two charts.}
 \label{fig:rf}
\end{figure*}

We assumed $R_{max}=7$, giving 8 levels corresponding to rating classes: Aaa, Aa, A, Bbb, Bb, B, C, and D. 

A simulation was run for $N=1000$ firms and was repeated 1000 times for different realizations of $J_{ij}$ distribution. For a chosen realization of $J_{ij}$, variable $s_i$ of a randomly selected firm $i$ was updated according to (\ref{eq:Ps}). Next, rating $R_i$ was updated according to (\ref{eq:R}).
 This was repeated $N$ times and was treated as one time step. After 8 time steps, the number of defaults ($ND$) were counted.

Although, in general, function $f(R,s)$ should lead to an empirical rating migration matrix \cite{Jafry,McNulty}, we studied two simplified cases.

\subsection{The case $f\equiv 0$}

The case $f\equiv 0$ means that the evolution of a firm's condition is only influenced by the environment and does not depend on the firm's present rating. In other words, neighbors aside, the probabilities of changes upward, downward, or  remaining at the same level are equal. Simulations for different $J_0$ (Figs. \ref{fig:rp}, \ref{fig:rf}) show that the distribution of $ND$ is different for small and large $J_0$.

If we denote a number of bankruptcies for a specific realization of $J_{ij}$ by $ND_k$, we can define a mean number of bankruptcies 
\begin{equation}
 \langle ND\rangle=\frac{1}{K}\sum_{k=1}^K ND_k,
\end{equation}
which is $ND_k$ averaged over $K$ different realizations. As a measure of risk, corresponding to unexpected losses, we can define an upper semivariance:
\begin{equation}
 Var_+=\frac{1}{K-1}\sum_{k=1}^K (ND_k - \langle ND \rangle)_+^2,
\end{equation}
where $(\cdot)_+$ is the Heaviside function. The semivariance measures the average square distance between the mean value $\langle ND \rangle$ and $ND_k>\langle ND\rangle$. It indicates the deviation of a statistical observation from the mean value in the pessimistic direction. Of course, it cannot be directly translated into confidence levels, as in the Gaussian case, but it can express the relative change of the risk.

\begin{figure}
 \includegraphics[angle=-90, scale=0.5]{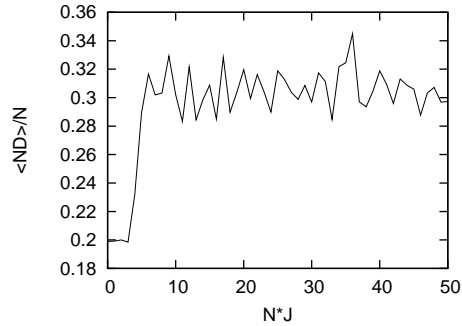}
 \caption{$ND$ as a function of $J_0$ for $f\equiv 0$ and $\sigma=0.001$}
 \label{fig:1}
\end{figure}

\begin{figure}
 \includegraphics[angle=-90, scale=0.5]{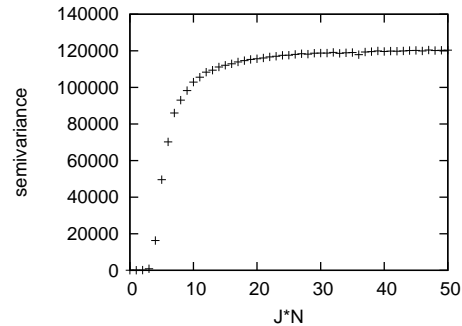}
 \caption{Semivariance as a function of $J_0$ for $f\equiv 0$.and $\sigma=0.001$}
 \label{fig:2}
\end{figure}

In order to examine the relation between $ND$ and $J_0$, and also between $Var_+$ and $J_0$, simulations were run for changing $J_0$ and fixed $\sigma_J=0.001$. The behavior of a mean number of defaults and their semivariance (Figs. \ref{fig:1}, \ref{fig:2}) signifies a phase transition. 

If one forgets the $R$ parameter, which in the case of $f\equiv 0$ does not influence the evolution of $s$, one gets a model of 3-state Potts glass. For such a model, three different phases can be observed, i. e., a paramagnetic, ferromagnetic, and spin-glass phase \cite{Nishimori}. Beyond a spin-glass phase (i. e., for  $\sigma_J<3/\sqrt{N}$), a mean field approach can be used to estimate the critical $J_c=3/N$ and explain the observed behavior. For a mean interaction parameter $J_0$ smaller than the critical $J_c$, the system of $N$ firms is in a paramagnetic phase. In this phase, a number of defaults is characterized by the distribution presented in Fig. \ref{fig:rp}, with a relatively low mean value and semivariance. For $J_0$ greater than $J_c$, the system is in a ferromagnetic phase, which is characterized by a jump in the mean number of defaults and semivariance. The jump is caused by a collective bankruptcies phenomenon  that can be observed in Fig. \ref{fig:rf} as a local maximum positioned around $ND=900$.

Results of our numerical simulations can be explained by a single analytical theory. With the use of a mean field approximation, we can estimate the probability $p$ that a firm will increase its rating ($s_i=1$) and $q$ that it will decrease ($s_i=-1$):
\begin{equation}
 p=\frac{\exp(J_0p)}{\exp(J_0p)+\exp(J_0q)+\exp(J_0(1-p-q))},
\end{equation}
\begin{equation}
 q=\frac{\exp(J_0q)}{\exp(J_0p)+\exp(J_0q)+\exp(J_0(1-p-q))},
\end{equation}

Because $s$ can take three values, we have: $p+q\leq 1$.

For an initially equally distributed portfolio, the mean number of defaults after time $t=8$ is given by:

\begin{equation}
\begin{split}
&\langle ND(p,q)\rangle= \frac{N}{7}[p\,{q}^{7}+\left( p-14\,{p}^{2}\right) \,{q}^{6}+\left( 12\,{p}^{2}+p\right) \,{q}^{5}\\
&+\left( 70\,{p}^{4}-80\,{p}^{3}+10\,{p}^{2}+p\right) \,{q}^{4}+ \\
&\left( 70\,{p}^{5}-120\,{p}^{4}+40\,{p}^{3}+8\,{p}^{2}+p\right) \,{q}^{3}\\
&+\left( 30\,{p}^{5}-60\,{p}^{4}+24\,{p}^{3}+6\,{p}^{2}+p\right) \,{q}^{2}+\\
&\left( -21\,{p}^{7}+80\,{p}^{6}-102\,{p}^{5}+32\,{p}^{4}+12\,{p}^{3}+4\,{p}^{2}+p\right) \,q\\
&-7\,{p}^{8}+21\,{p}^{7}-24\,{p}^{6}+2\,{p}^{5}+8\,{p}^{4}+4\,{p}^{3}+2\,{p}^{2}+p].
\end{split}
\end{equation}

In a paramagnetic phase, $p=q=1/3$, so $\langle ND(1/3,1/3) \rangle$ $\approx 0.202$. In a ferromagnetic phase, a system can get ordered in three ways, so different $(p,q)$ are possible: (0,0), (1,0), (0,1). Due to symmetry, all of them are equally probable. Hence, the mean number of defaults is equal to:
\begin{equation}
\begin{split}
  \langle ND\rangle=&\frac{1}{3}(\langle ND(0,0)\rangle+\langle ND(1,0)\rangle+\langle ND(0,1)\rangle).
\end{split}
\end{equation}
In this case $\langle ND\rangle=N/3$.
\begin{figure}
 \includegraphics[scale=0.5, angle=-90]{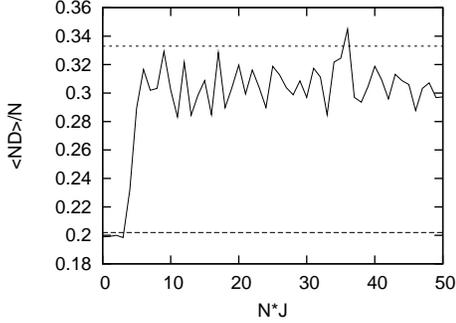}
 \caption{The mean number of defaults (ND) and levels of ND corresponding to the two phases ($ND/N=0.202$, $ND/N=0.305$).}
 \label{fig:levels}
\end{figure}
Transition between the two phases is ilustrated in Fig. \ref{fig:levels}. The mean number of defaults jumps from the level $ND/N=0.202$ to the upper level, which is lower than the predicted $0.333$. This discrepancy is caused by the value of $q$ being different from 1 in the beginning while the system is being ordered. The predicted value is reached in a longer simulation time.

\begin{figure}
\includegraphics[angle =-90, scale=0.5]{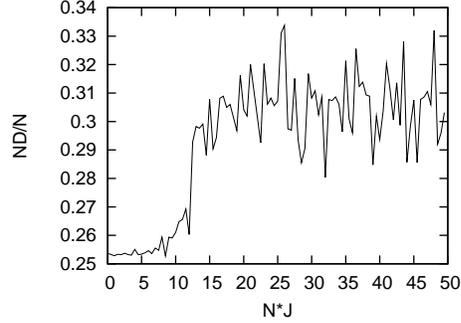}
\caption{$ND$ as a function of $J_0$ for $f\equiv 0$ and $\sigma_J=0.2$}
\label{fig:5a}
\end{figure}

\begin{figure}
\includegraphics[angle =-90, scale=0.5]{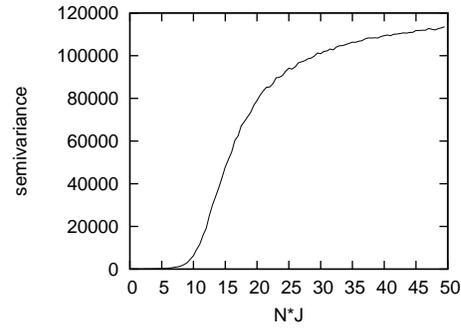}
\caption{Semivariance as a function of $J_0$ for $f\equiv 0$ and $\sigma_J=0.2$}
\label{fig:5b}
\end{figure}

A spin-glass phase is characterized by  a nonvanishing spin-glass order parameter \cite{Nishimori2}. Considering only magnetization, the spin-glass phase is indistinguishable from the paramagnetic, which is why the bankruptcy mechanism looks similar in both phases. The transition from spin-glass to ferromagnetic phase is visualized in Figs. \ref{fig:5a}, \ref{fig:5b}.

The appearance of different phases is very important from the systemic risk point of view. The paramagnetic phase is characterized by a moderate number of relatively low correlated defaults. In the ferromagnetic phase, the evolution can end up in one of two possible scenarios: (\textit{i}) a low number of defaults, firms are in good shape and strong interactions prevents them from going into default and (\textit{ii}) a very high number of defaults, bankrupting firms pull their partners down.  The existence of these two scenarios makes the average $\langle ND\rangle$ differ very little in magnitude, compared to paramagnetic $\langle ND\rangle$. In spite of this, risk of collective failure grows significantly, which is signaled by the behavior of the semivariance.

\subsection{The case of nonzero $f$}
The case of a constant nonzero $f$ that is independent of $R$ is also a simplification. However, it is more realistic than $f\equiv 0$, because it relates the probability of rating change to the direction of the change.

 We assume here that $f(R, s)$ depends only on $s$ such that: $\exp(f(R,-1))=0.15$, $\exp(f(R,0))=0.75$, and  $\exp(f(R,1))=0.10$. A firm that does not interact with neighbours, therefore, has a very high probability of staying at the present rating level and a smaller probability of changing the level. Furthermore, the probability that the firm goes down is higher than that it goes up. This assumption is based on the observation of empirical transition rates between rating classes \cite{Landschoot}. 

\begin{figure}
 \includegraphics[angle=-90, scale=0.5]{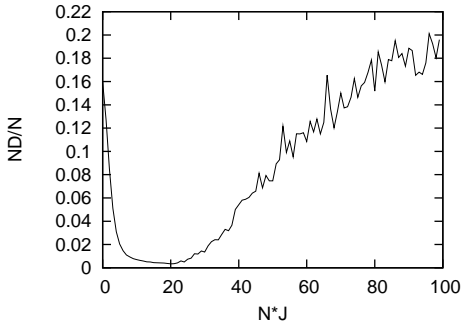}
\caption{$ND$ as a function of $J_0$ for a constant field.}
 \label{fig:3}
\end{figure}

\begin{figure}
 \includegraphics[angle=-90, scale=0.5]{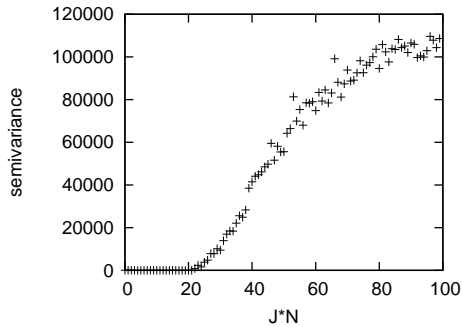}
\caption{Semivariance as a function of $J_0$ for a constant field.}
 \label{fig:4}
\end{figure}

The probability of bankruptcy as a function of $J_0$ is presented in Fig. \ref{fig:3}. It has a minimum for $J_0N\approx 20$. Semivariance jumps abruptly for the same value (Fig. \ref{fig:4}). As far as $ND$ is concerned, there is an optimal value of $J_0$ that minimizes the number of defaults. Increasing the value of mean interaction strength $J_0$, stabilizes the flipping spins in the beginning. But when the interactions get stronger and the system is in the ferromagnetic phase, the collective bankruptcy phenomenon increases the mean number of bankruptcies. However, from the risk point of view, $J_0N$ should not exceed the critical value, because it makes the semivariance jump abruptly.

Introduction of nonzero $f$ which is equivalent to breaking symmetry by a kind of external field in a Potts glass model, shifts the critical $J$ ($J<20$) and changes the shape of ND distribution into one with a minimum. The parameter $f$ provides a stabilizing force for small $J$ and reduces the number of bankrupts. For a large $J$ this force is overwhelmed by collective bankruptcies in the ferromagnetic phase.

The effect of a minimal number of bankruptcies for an average magnitude of $J_0$ agrees with the results of Lorentz and Battiston \cite{Lorenz}. They investigated a model of financially connected firms, showing that the number of bankruptcies in the network is minimized for an intermediate density of links.

\section{Conclusions}

We proposed a simple model of bankrupting firms where we allowed companies to change their ratings under the influence of two factors. The first was individual dynamics with one-step memory making the probability of rating changes depend on the present rating.  The second factor included mutual interactions  between firms. The interaction matrix was chosen from Gaussian distribution, characterized by parameters $J_0$ and $\sigma_J$. 

We simulated bankruptcies in our model for two simplified cases. We observed phase transitions, which can be described with the use of the theory of Potts spin-glass. 

For a credit portfolio, it is obvious that unexpected losses grow with rising correlations. However, our study shows that, considering an economic interaction, one can expect an abrupt jump of unexpected losses as a function of interaction strength due to phase transition.

Portfolio expected losses are calculated as the mean value of losses. The correlations do not affect expected losses.  Our analysis shows that the mean number of defaults depends on the interaction strength. In the case of $f\equiv 0$, it is when the evolution of a firm's condition is only influenced by the environment and does not depend on the present rating of a firm, the mean number of defaults jumps. More interesting is the analyzed case of nonzero $f$. If $J$ rises, then for small values it stabilizes the system, since the function $f(R,s)$ prefers ordering in a neutral ($s=0$) state. For high values of $J$, the stabilizing role of the factor $f(R,s)$ can be neglected as compared to inter-firm interactions. As a result, large collective bankruptcies occur in the ferromagnetic phase. 

The relation of growing $ND$ with growing mean interaction strength $J_0$ may seem to contradict empirical evidence about competition and default probability. Eisdorfer and Hsu \cite{Eisdorfer} have shown that, considering technology-intensive industries, there is a strong positive relationship between the level of competition and the number of bankruptcies.  They measure the level of competition by a number of patents in a sector and show that high competition leads to higher frequency of bankruptcy when poorly performing firms are more likely to bankrupt. In the real economy, competition induces a higher default rate among small and beginning firms and conserves a dominating position of big and well-off companies. Since there is a larger number of small firms compared to large firms, the default rate is higher for higher competition level. For example, patent competition in a software market is very painful for smaller, especially starting, firms and is very convenient for a large software corporation. A positive corelation between competition level and a number of bankruptcies is obvious in this case.  

In the simulations of our model, we considered a symmetric interaction matrix with values from a Gaussian distribution. As an initial state, we took uniformly distributed variables $R_i$ and $s_i$. All firms in the model have an influence on their neighbours of the same strength. To adopt the model to the situation described by \cite{Eisdorfer}, one needs to modify it. The modification could include the following points:
\begin{itemize}
 \item Interactions are asymmetric. The influence of a bigger firm on a smaller one should be greater than the reverse.
\item Bankrupted nodes are replaced by new ones.
\item In general, new nodes are smaller than older nodes and they start with a lower rating grade.
\item There are more small firms than big ones. 
\end{itemize}
For such a modified model, starting with an empirical distribution of firms' sizes and ratings, an effect of positive dependence between competition strength and number of bankruptcies should be observable. It would also be interesting to implement the evolution of the sizes of nodes and interplay between this evolution and other model parameters, as well as to choose function $f(R,s)$ closer to empirical data.

\section*{Acknowledgements}
This project was supported by a special grant from Warsaw University of Technology.

\end{document}